\documentclass{svproc}
\usepackage{url}
\usepackage{amsmath}
\usepackage{float}
\usepackage{graphicx}

\begin{document}
\mainmatter              
\title{Vibration-Based Bridge Health Monitoring using Telecommunication Cables}
\titlerunning{Vibration-Based BHM using Telecommunication Cables}  
%
\author{Jingxiao Liu \and Siyuan Yuan\and 
Bin Luo \and Biondo Biondi \and Hae Young Noh}
\authorrunning{Liu et al.} 
%
\tocauthor{Jingxiao Liu, Siyuan Yuan, Bin Luo, Biondo Biondi, Hae Young Noh}
\institute{Stanford University, Stanford CA 94305, USA}

\maketitle              

\begin{abstract}
Bridge Health Monitoring (BHM) enables early damage detection of bridges and is thus critical for avoiding more severe damages that might result in major financial and human losses. However, conventional BHM systems require dedicated sensors on bridges, which is costly to install and maintain and hard to scale up. To overcome this challenge, we introduce a new system that uses existing telecommunication cables for Distributed Acoustic Sensing (DAS) to collect bridge dynamic strain responses. In addition, we develop a two-module physics-guided system identification method to extract bridge damage-sensitive information (e.g., natural frequencies and mode shapes) from noisy DAS data by constraining strain and displacement mode shapes by bridge dynamics. This approach does not require installation and maintenance of dedicated sensors on bridges. We evaluate our system with field experiments on a concrete bridge with fiber cable running in a conduit under the deck. Our system successfully identified modal frequencies and reconstructed meter-scale mode shapes.
\end{abstract}
\section{Introduction}
Aging bridges pose challenges to the economy and public safety. According to the National Bridge Inventory, around 140,000, or 22\% of more than 617,000 bridges in the United States are considered structurally deficient or functionally obsolete in 2019~\cite{asce}. The state of
aging bridges demands researchers to develop efficient and effective bridge health monitoring (BHM) approaches for keeping our bridges safe and reliable.

Currently, BHM is mainly based on manual inspection~\cite{sony2019}, which is labor-intensive, time-consuming, and potentially hazardous. To address such drawbacks, dedicated sensor-based BHM approaches, where bridges are instrumented with sensing devices, were introduced to autonomously collect structural performance data; however, these approaches are also costly and inefficient as they require on-site installation and maintenance of dedicated sensors on every bridge. To improve efficiency, many researchers recently proposed mobile sensing methods for BHM, for example, capturing dynamic information by scanning the bridge using vehicles in a non-dedicated manner~\cite{liu2020,liu2021}. Yet, such mobile sensing methods sacrifice temporal information to get high-spatial-resolution information of multiple bridges, restricting their ability to diagnose bridge conditions continuously.

To this end, we introduce a novel BHM system that uses existing telecommunication cables as non-dedicated distributed acoustic sensors to collect bridge dynamic responses. There are millions of kilometers of optical fiber cables deployed around the world. These extensively installed telecommunication cables are transformed into distributed acoustic sensing (DAS) systems, which are long-offset virtual and dynamic strain sensor arrays with a high temporal and spatial resolution~\cite{yuan2021}. DAS responses collected from cables bounded to bridges reflect bridge dynamics and contain damage-sensitive information. Such information can be used for detecting and diagnosing bridge damages. In addition, one individual DAS interrogator of the latest generation can record data from a telecommunication cable up to 100 km long that would cover multiple bridges. Thus, it is a cost-effective way to perform a large-scale BHM using DAS responses from telecommunication cables as it will not cause interference to other signals nor require the installation and maintenance of dedicated sensors.

In this work, we extract three damage-sensitive features using telecommunication responses, namely bridge natural frequencies, strain mode shapes (SMSs), and displacement mode shapes (DMSs). Natural frequencies and DMSs are sensitive to bridge global behavior, while SMSs are sensitive to bridge local behavior~\cite{hong2015}. However, estimating these damage-sensitive information is challenging due to the higher noise level from telecommunication DAS than from other dedicated sensors. Especially, error would propagate from SMSs' estimation using noisy DAS responses to DMSs' estimation as it requires one to double integrate the estimated SMSs either numerically or analytically. As the noise propagates, numerical double integration of SMS becomes unstable. Analytical double integration of SMS with conventional basis functions, such as polynomial functions, will produce inaccurate results~\cite{thong2004}.

To overcome the challenge, we introduce a two-module physics-guided method that estimates bridge damage-sensitive features. In the first module, we formulate an elemental strain state-space model that considers telecommunication cable responses as observations of bridge strain dynamics. This model represents the DAS system in a more compact and convenient way than other representations of dynamical systems. We then estimate the bridge natural frequencies and SMSs using the data-driven stochastic subspace identification (SSI-data) algorithm~\cite{van2012}. SSI-data is used because it does not need optimization techniques required by other methods such as prediction error minimization~\cite{reynders2012}. In the second module, we derive a parametric function based on bridge dynamics and fit the estimated SMS to the parametric shape function. Then, the fitted function is double integrated for estimating the DMS. This two-module method addresses the unstable problem encountered by numerical double integration methods and improves DMSs’ estimation accuracy by physically constraining the SMSs. 

Our method was validated with field experiments on a concrete bridge in San Jose, California, with telecommunication fiber cables running in a conduit beneath the deck. Our system identified the first three natural frequencies with a 0.06 Hz mean absolute difference (MAD) compared to those extracted from accelerometers deployed on the bridge deck. In addition, our system estimated the first three SMSs and DMSs in a meter-scale resolution and achieved a 0.80 modal assurance criterion (MAC) compared to those from accelerometers.

\section{Vibration-based BHM system using DAS with telecommunication cables}

\begin{figure}[htb]
    \centering
    \begin{minipage}{0.9\textwidth}
        \centering
        \includegraphics[width=1\linewidth]{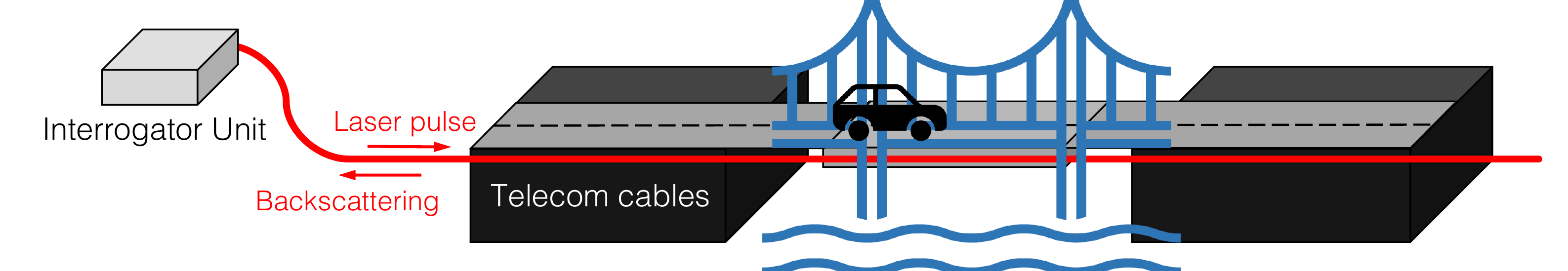}
    \end{minipage}
    \quad
    \begin{minipage}{1\textwidth}
        \centering
        \includegraphics[width=1\linewidth]{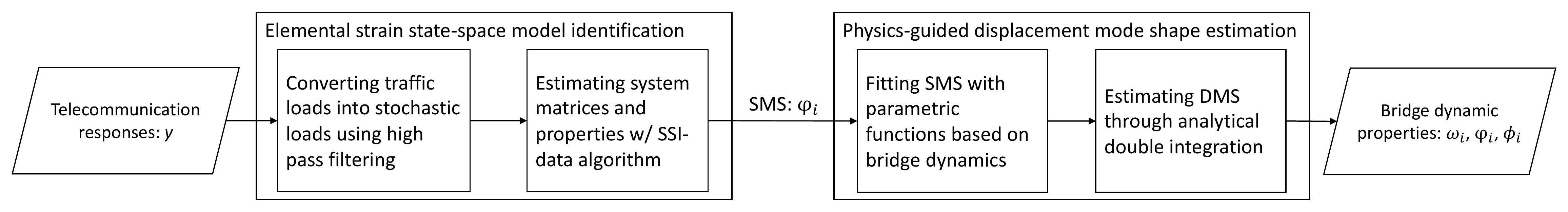}
    \end{minipage}%
    \caption{Our BHM system using DAS with existing telecommunication cables.}
    \label{fig:system}
\end{figure}

This section introduces our BHM system (as shown in Figure~\ref{fig:system}) which uses DAS response from telecommunication cables. The method contains two modules: 1) an elemental strain state-space model identification module and 2) a physics-guided displacement mode shape estimation module.

\subsection{Elemental strain state-space model identification}

This module has two steps. In the first step, a high-pass filter with a 1-Hz cutoff frequency is applied to convert unmeasured loads into stochastic dynamic loads and exclude the low-frequency loads due to traffic traveling, which enables us to identify the system without measuring the input loads. Then, we formulate our problem as an elemental strain state-space model that considers DAS responses as observations, $y(t)$, of the actual bridge strain dynamics, $x(t)$:

\begin{equation}
\begin{aligned}
    &\dot{x}(t)=Ax(t)+w(t)\\
    &y(t)=Cx(t)+v(t)
\end{aligned}
\label{eq:sss}
\end{equation}

where $A$ and $C$ are the system and observation matrices, respectively; $w(t)$ and $v(t)$ are the stochastic load vector and observation noise, respectively.
In the second step, the SSI-data algorithm is adopted to estimate the system matrix and bridge properties. This step outputs the estimated bridge natural frequencies, $\omega_i$, and SMSs, $\varphi_i$ for different mode $i$. SMSs are then input to the next module for estimating DMSs.

\subsection{Physics-guided displacement mode shape estimation}

According to the beam theory, computing the DMS, $\phi_i$, requires double integration of the SMS. This double integration can be performed numerically or analytically. However, due to large measurement noise from telecommunication cables, conventional numerical and analytical double integration methods (e.g., the trapezoidal rule and analytical integration with a polynomial basis) can be inaccurate as the noise propagates in integration steps.
To this end, we estimate the DMS by first fitting the estimated SMS function with a physics-guided parametric function and then double integrating the fitted SMS. The physics-guided shape function is obtained by using the homogeneous solution for beam vibrations. For a continuous beam, the general solution for the $i$-th mode’s SMS of the $k$-th beam span is:
\begin{equation}
\begin{aligned}
    \varphi_{i,k}(l)=&-d\beta_{i,k}^2(-C_{1,i,k}\sin(\beta_{i,k} l)-C_{2,i,k}\cos(\beta_{i,k} l)\\
    &+C_{3,i,k}\sinh(\beta_{i,k} l)+C_{4,i,k}\cosh(\beta_{i,k} l)),\quad\quad \text{for~}0\leq l\leq l_k
    \label{eq:strain}
\end{aligned}
\end{equation}
where $\beta_{i,k}$ depends on beam properties; $d$ is the distance from strain measurement point to the beam neutral axis; $l_k$ is the length of the $k$-th beam span; $C_{1,i,k}$, $C_{2,i,k}$, $C_{3,i,k}$, and $C_{4,i,k}$ are constant variables that are evaluated from boundary conditions, including zero displacement, continuous angle of rotation and bending-moment at support locations. In this way, we address the unstable problem caused by numerical double integration and improve the estimation accuracy of analytical double integration by physically constraining the SMS.

\section{Field evaluation}
This section presents the evaluation of our system with field experiments on a 50-meter-long concrete continuous span bridge in San Jose, California.

\subsection{Experimental setup and data description}
\begin{figure}[htb]
    \centering
    \begin{minipage}{0.35\textwidth}
        \centering
        \includegraphics[height=2.5cm]{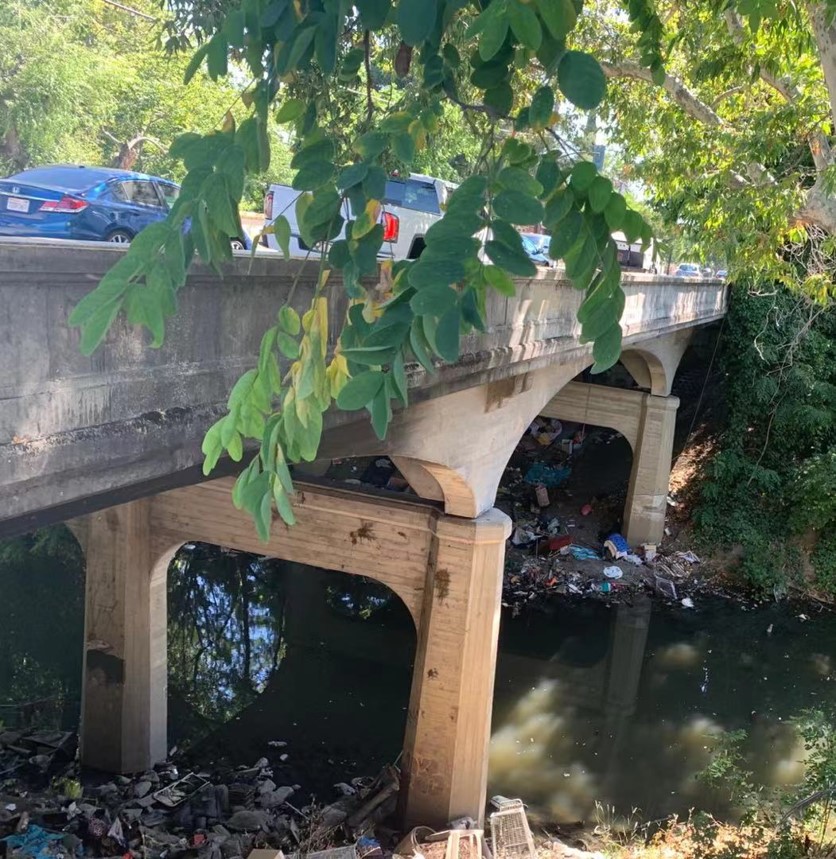}
        
        {(a) The Coyote Creek bridge}
    \end{minipage}
    \quad
    \begin{minipage}{0.6\textwidth}
        \centering
        \includegraphics[height=2.5cm]{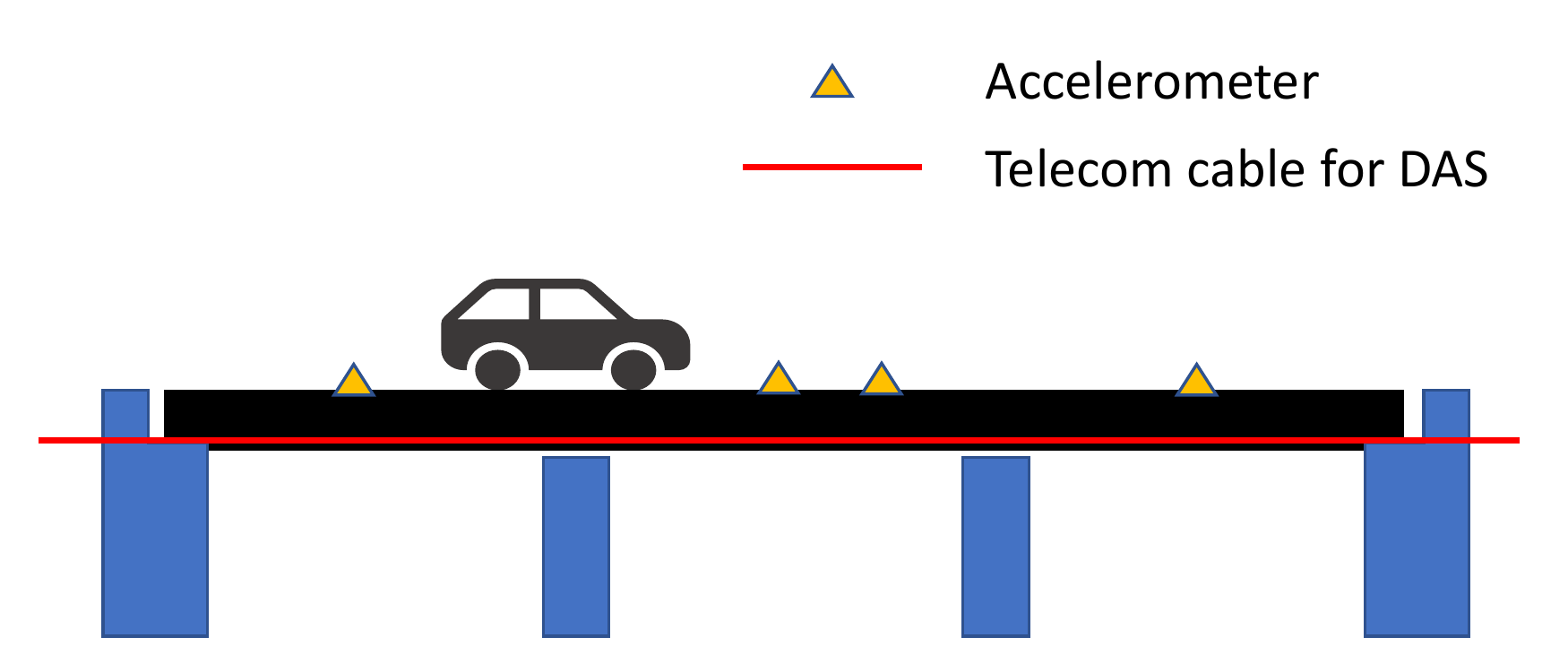}
        
        {(b) A representation of our sensing deployment}
    \end{minipage}%
    \caption{Our experimental setup: (a) the testbed bridge and (b) our sensing deployment.}
    \label{fig:evaluation}
\end{figure}

Figure~\ref{fig:evaluation} shows the testbed bridge (the Coyote Creek bridge) and a representation of our sensing deployment. The Coyote Creek bridge is a concrete girder continuous three-span bridge that carries highway and pedestrian walkway over the Coyote Creek in San Jose, California. Telecommunication fiber cables run in a conduit beneath the bridge deck. An interrogator that was installed around 2 kilometers away from the bridge performed distributed acoustic sensing with a one-meter channel spacing and 250 Hz sampling rate. Furthermore, four accelerometers were installed on top of the bridge deck to measure the vertical acceleration of the bridge, providing validation signals to the DAS system. Specifically, three of them were installed in the middle of each span, and the last one was installed at the third quarter of the middle span. During the experiment, we collected ten sets of DAS and bridge acceleration data on two consecutive days in June 2021. Each set contains around eight minutes of data.

\subsection{Results and discussions}
\begin{figure}[tbh]
    \centering
    \includegraphics[width=1\linewidth]{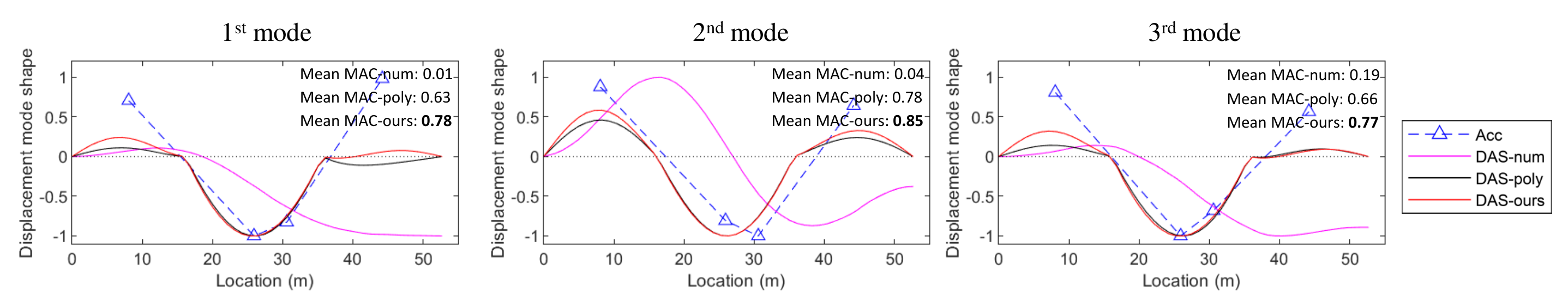}
    \caption{Estimated DMSs using numerical double integration (num), polynomial-based analytical double integration (poly), and our method (ours) and their Mean MAC using accelerometer data (Acc) vs. DAS data.}
    \label{fig:mac}
\end{figure}

Table~\ref{tab:freq} presents mean values of identified bridge natural frequencies using the ten-set accelerometer and DAS data and mean absolute difference (MAD) between the two systems. Our system identified the first three natural frequencies with a 0.06 Hz MAD compared with those identified from accelerometer data. In addition, our system effectively estimated the first three SMSs and DMSs. We compare the estimated DMSs using our method with those using two baseline methods, including a numerical method based on the trapezoidal rule and an analytical double integration with a polynomial basis. Figure~\ref{fig:mac} shows the estimated DMSs and their mean MAC values using accelerometer versus using DAS data for different double integration methods. Our method achieves 0.80 mean MAC for the first three modes, which has a 72\% and an 11\% improvement compared to the numerical method and the polynomial-based method, respectively.

\begin{table}[htb]
  \centering
  \caption{Mean values of identified bridge natural frequencies using accelerometer data (Freq-acc) and DAS data (Freq-DAS), and the MAD between them.}
  \begin{tabular}{cccc}
    \hline
    Mode \# & Mean freq-acc (Hz)& Mean freq-DAS (Hz)& MAD (Hz)\\
    \hline
    1& 4.57& 4.61&0.03\\
    2& 6.33& 6.32& 0.03\\
    3& 8.94& 8.85& 0.11\\
    \hline
  \end{tabular}
  \label{tab:freq}
\end{table}

\section{Conclusions}
This paper introduces a non-dedicated bridge health monitoring system that turns existing telecommunication cables into distributed acoustic sensors to capture bridge dynamic responses. This system is efficient and cost-effective as it does not require on-site installation and maintenance of sensors and equipment by taking advantage of extensively installed telecommunication cables. To overcome the large measurement noise challenge, we developed an elemental strain state-space model followed by a physics-guided analytical double integration to estimate damage-sensitive dynamic properties of bridges. We evaluated our system with a concrete three-span bridge and validated our estimations with conventional accelerometers. Our system identified the first three natural frequencies, strain mode shapes, and displacement mode shapes of the bridge, outperforming two baseline methods that use conventional numerical and analytical double integration methods.

%

\end{document}